\documentclass[floats,floatfix,showpacs,amssymb,prd,twocolumn,superscriptaddress,nofootinbib]{revtex4-1}

\usepackage{amssymb,amsmath,verbatim,mathtools,needspace,etoolbox,graphicx,physics,microtype,afterpage,bm}
\usepackage[dvipsnames, usenames]{xcolor}
\usepackage[english]{babel}
\usepackage{lmodern}
\usepackage{booktabs}
\usepackage{amsfonts}
\usepackage{listings}
\usepackage{fancyhdr}
\usepackage{xcolor}
\usepackage{comment}
\usepackage{enumerate}

\graphicspath{{./figures/}}

\definecolor{rb4}{HTML}{27408B}

\def\Tobs{T}

\def\D90{\mathcal{D}^{90\%}}
\def\pdet{p_\mathrm{det}}
\def\pfa{p_\mathrm{fa}}
\def\depth{\mathcal{D}}
\def\depthmf{\mathcal{D}_\mathrm{MF}^{90\%}}

\newif\ifshowfigs
\showfigstrue

\definecolor{linkcolor}{rgb}{0.0,0.3,0.5}
\usepackage[unicode, colorlinks=true, linkcolor=linkcolor, citecolor=linkcolor, filecolor=linkcolor,urlcolor=linkcolor, pdfusetitle]{hyperref}
\usepackage[all]{hypcap}
\usepackage[T1]{fontenc}
\usepackage[utf8]{inputenc}
\usepackage{tabularx}
\usepackage{float}
\allowdisplaybreaks
\begin{document}

\title{Attention U-Net for all-sky continuous gravitational wave searches}

\author{Damon H. T. Cheung}
\email{damoncht@umich.edu}
\affiliation{
Department of Physics, University of Michigan, 450 Church St, Ann Arbor, MI 48109
}

\date{\today} 

\begin{abstract}
Detecting continuous gravitational waves is challenging due to the high computational cost of template-based searches across large parameter spaces, particularly for all-sky searches. 
Machine learning offers a promising solution to perform these searches with reasonable computational resources. 
In this study, we trained an attention U-Net, a convolutional neural network, on $\approx 10.67$ days simulated data with Gaussian noise for all-sky searches at different frequencies within the 20--1000\,Hz band.
Our model trained at 20 Hz achieves the best sensitivity, with a 90\% detection efficiency sensitivity depth \(\D90=29.97 \pm 0.24\,\mathrm{Hz}^{-1/2}\) with 1\% false alarm rate per 50mHz, while the model trained on the entire 20--1000\,Hz band yields \(\D90=18.63 \pm 0.24\,\mathrm{Hz}^{-1/2}\).
The sensitivities achieved are comparable to state-of-the-art results using deep learning approaches, with less than 50\% of the training time and data.
We find that sensitivity scales as $\Tobs^{0.28\pm 0.01}$ with total observation time for the attention U-Net trained at 20\,Hz, similar to semi-coherent search methods.
The neural network demonstrates robustness on datasets with time gaps, with sensitivity dependence on duty factor analyzed.
We also investigated the sensitivity dependence of the trained attention U-Net models on sky location. 
Our findings show that attention U-Net is a scalable and effective approach for all-sky continuous gravitational waves searches.
\end{abstract}
\maketitle

\section{Introduction}
Continuous gravitational waves (CWs) are faint, long-duration, quasi-monochromatic signals emitted by fast spinning, non-axisymmetric neutron stars (NSs), which have not yet been detected. 
Detecting these signals offers a unique opportunity to study NS properties and the physics of dense matter, a key unsolved problem in astrophysics~\citep{Riles2023}.

The amplitude of CWs is expected to be significantly weaker than the noise amplitude of current gravitational wave detectors. To achieve sufficient sensitivity, integrating a long duration of data is required to increase the signal-to-noise ratio (SNR). 
However, the most sensitive method, template-based coherent matched filtering, demands an increasing number of templates for longer observation times, with computational cost scaling as $\sim \Tobs^n$, where $n \sim 5$ to 6 or higher. 
This limits coherent searches to a few weeks of data. Alternatively, semi-coherent methods divide the total observation time into shorter segments, coherently analyzing each segment and combining the segment detection statistics incoherently. 
Despite their reduced computational burden, semi-coherent methods remain challenging and computationally bound for all-sky CW searches because of the vast parameter space involved. 
To overcome these computational limitations, machine learning offers a promising approach for effectively navigating the extensive parameter space.

Several studies have explored machine learning for CW detection. 
For instance, \citep{Miller2019, Modafferi2023, Attadio2024} employed convolutional neural networks (CNNs) to detect long-duration transient CWs.
Other works used CNNs to accelerate the follow-up process \citep{Morawski2020, Beheshtipour2020, Beheshtipour2021, Takahiro2021} and mitigate instrumental artifacts \citep{Bayley2020}.
Additionally, \citep{Bayley2022} utilized a conditional variational autoencoder to expedite parameter estimation post-search. 

For all-sky CW searches, all models address signal detection as a classification problem. 
\citep{Dreissigacker2019, Dreissigacker2020} trained a ResNet on $\approx11.57$ days of data with Gaussian noise. 
\citep{Yamamoto2022} used a CNN to detect signals with Gaussian and sinusoidal line noise.
\citep{Joshi2023, Joshi2024} developed a CNN with residual blocks to look for the presence of signal on 10 days of data.
\citep{Joshi2025} trained a transformer-based model on 1 day of data.
These represent the only end-to-end machine learning-based all-sky search models to date. Among them, \citep{Joshi2024} achieved the highest sensitivity across the largest parameter space explored so far.

In this work, we employ an attention U-Net~\citep{Oktay2018}, a CNN commonly used for image segmentation, to develop an end-to-end all-sky CW search model. 
We treat the input data as images and train the U-Net to map noisy inputs to denoised signal outputs, using the mean power of the output image as the detection statistic.
Five models were trained for all-sky CW signals at 20, 200, 500, 1000, and the entire 20--1000\,Hz band, respectively. 
Our model trained at 20\,Hz achieves the best sensitivity, with a sensitivity depth (see Sect.~\ref{sec:data}) of \(\D90 = 29.97 \pm 0.24\,\mathrm{Hz}^{-1/2}\) at 90\% detection efficiency, and \(\D90 = 18.63 \pm 0.24\,\mathrm{Hz}^{-1/2}\) for the model trained on the entire 20--1000\,Hz band.
The sensitivities are comparable to the results reported by~\citep{Joshi2024}, using less than 50\% of the training time and data, requiring approximately 5 days for training.
This serves as a proof of concept that the attention U-Net is a suitable and effective approach for CW searches.
For the 20\,Hz model, we extended the observation time, achieving a sensitivity depth of \(\D90 = 35.86 \pm 0.38\,\mathrm{Hz}^{-1/2}\) with $\Tobs \approx 21$ days, with sensitivity scaling as $\Tobs^{0.28 \pm 0.01}$, comparable to the $\Tobs^{0.25}$ scaling of semi-coherent searches~\citep{Wette2012}. 
The training time scales linearly with observation time, more efficient when compare the to semi-coherent method $\sim \Tobs^2$ dependece~\citep{Prix2012}.
When tested on datasets with time gaps, the model, trained on continuous data, exhibits robustness under real data conditions, with sensitivity dependence on duty factor scaling as $\eta^{0.53 \pm 0.02}$.
Additionally, we analyzed the sensitivity dependence of the trained models on sky location to investigate their limitations.

This article is organized as follows. 
Section~\ref{sec:data} outlines the CW waveform models. 
Section~\ref{sec:method} describes the attention U-Net architecture, training process, and detection statistic for assessing detectability.
Section~\ref{sec:result} evaluates the performance of the trained attention U-Net models on test data. 
Finally, Section~\ref{sec:discussions} explores potential improvements and future research directions.

\section{Waveform Models} \label{sec:data}
To train our neural network, we prepare data modeling CW signals embedded in a noise background. 
The data collected from GW detectors, denoted as $d(t)$, consist of a time series expressed as
\begin{equation}
    d(t) = n(t) + h(t),
\end{equation}
where $n(t)$ represents the detector noise and $h(t)$ denotes the CW signal.
The CW signal from an isolated, rapidly spinning, non-axisymmetric NS is given by
\begin{align}
    h(t) = h_0 \bigg[ F_+(t, \alpha, \delta, \psi) \frac{1 + \cos^2 \iota}{2} \cos \Phi(t) \notag \\
    + F_\times(t, \alpha, \delta, \psi) \cos \iota \sin \Phi(t) \bigg],
    \label{eq:snr_strain_eqn}
\end{align}
where $t$ is the time in the detector frame, $\alpha$ and $\delta$ are the source’s sky location coordinates, $\psi$ is the polarization angle, $F_+$ and $F_\times$ represent the detector’s sensitivity to the sky location, and $\iota$ is the angle between the source’s spin axis and the line of sight to the detector. 
The phase of the CW signal is expressed as a Taylor series:
\begin{align} \label{eq:phase}
    \Phi(\tau) = \Phi_0 + f (\tau - \tau_0) + \frac{1}{2} \dot{f} (\tau - \tau_0)^2 + \cdots,
\end{align}
where $f$ is the Solar System
Barycenter (SSB) signal frequency, $\dot{f}$ is the first frequency derivative, $\tau$ is the
SSB time, $\Phi_0$ is initial phase, and $\tau_0$ is the reference time. 
The parameters in Eq.~\eqref{eq:snr_strain_eqn} govern the signal’s amplitude over time, while those in Eq.~\eqref{eq:phase} determine its trajectory in the spectrogram.

For CW data analysis, we convert the time series into the Short Fourier Transform (SFT) format by dividing the data into segments. 
Under the assumption of a slowly varying CW signal, SFT data enable signal detection while significantly reducing the size of the dataset.  
We use \textit{PyFstat}~\citep{pyfstat2021} to generate pure CW signals in SFT format, segmenting the data into time spans of $T_{\text{SFT}} = 144{,}000$\,s (4 hours) per segment. 
In practice, detector noise often exhibits non-Gaussian characteristics; however, for simplicity, we model it as Gaussian noise.
This noise is generated directly in the Fourier domain, with the real and imaginary components of the Fourier transform defined by
\begin{equation} \label{eq:Fourier_Gaussian_noise}
    \Re \{ \tilde{x}[t; f] \}, \Im \{ \tilde{x}[t; f] \} \sim \mathcal{N} \left( 0, \frac{1}{2} \sqrt{T_{\text{SFT}} S_n} \right),
\end{equation}
where $\Re \{ \tilde{x}[t; f] \}$ and $\Im \{ \tilde{x}[t; f] \}$ are the real and imaginary parts of the noise in SFT format $\tilde{x}[t; f]$ at timestamp $t$ and frequency $f$, $\mathcal{N}(\mu, \sigma)$ is a Gaussian distribution centered at $\mu$ with standard deviation $\sigma$, and $\sqrt{S_n}$ is the single-sided noise amplitude spectral density.

We use $\sqrt{S_n}/h_0$ to represent the noise level of the injected signal in our training and test data, as the ability to detect the signal depends on its amplitude relative to the detector's noise amplitude. 
In this work, we consider only the first-order spindown term in Eq.~\eqref{eq:phase}. The parameter space defining the signals is listed in Table~\ref{tab:parameter_ranges}.
We assess the neural network’s performance for an all-sky search with an observation time of $\Tobs = 921{,}600$\,s ($\approx$ 10.67 days), chosen to approximate 10 days and to yield $2^n$ time segments to favor the architecture of the neural network.

\begin{table}[h]
\centering
\begin{tabular}{lc}
\hline\hline
Start time           & 1 368 970 000 s \\
Duration             & 921 600 s ($\approx 10.67$ days)\\
Reference time       & 1 369 430 800 s \\
Frequency            & $f \in [20, 1000]\; \mathrm{Hz}$\\
Spin-down            & $\dot{f} \in [-10^{-10}, 0] \; \mathrm{Hz/s}$ \\
Inclination angle    & $\cos \iota\in[-1, 1]$ \\
Initial phase        & $\phi \in [0, 2\pi]$ \\
Polarization angle   & $\psi \in [-\pi/4, \pi/4]$ \\
Right ascension (RA) & $\alpha \in [0, 2\pi]$ \\
Declination (Dec)    & $\sin \delta\in[-1, 1]$ \\ \hline\hline
\end{tabular}
\caption{Parameters used to define the continuous wave signals for all-sky search.}
\label{tab:parameter_ranges}
\end{table}

\section{Methods} \label{sec:method}
In this section, we describe the architecture of the attention U-Net, the preprocessing of input data, loss function, the training process, and the detection statistic used to evaluate detectability.

Unlike~\citep{Dreissigacker2020, Joshi2024}, which use neural networks to output a single probability indicating the presence of a signal, we treat the input data as images and employ the attention U-Net~\citep{Oktay2018} to learn a mapping from noisy input images to clean, denoised, signal-only outputs, preserving the input dimensions.
Then, we compute a simple detection statistic on the output image.
Therefore, we have to prepare a set of noisy signal and clean signal image pairs as our training data.

Similar to~\citep{Dreissigacker2020, Joshi2024}, we use the real and imaginary parts of SFT data as a two-dimensional input. 
Combining data from the H1 and L1 detectors, we obtain four channels: real and imaginary components for each detector. 
We perform a Fourier transform every 14{,}400\,s on data spanning a total observation time of $\Tobs = 921{,}600$\,s, yielding 64 time segments along the time axis. 
The frequency axis is cropped to retain 512 frequency bins, corresponding to a bandwidth of 35.56\,mHz, sufficient to accommodate Doppler modulations due to Earth's motion over the observation period. 
To enhance robustness, we then randomly shift the frequency axis, ensuring the CW signal remains within the selected range without always appearing at the center. 
The resulting preprocessed data have a shape of $(512, 64, 4)$, with dimensions selected as powers of 2 to favor the U-Net’s architecture.
Finally, we normalize the data to the range $[-1, 1]$.

\subsection{Attention U-Net Architecture}

The attention U-Net~\citep{Oktay2018}, a convolutional neural network (CNN) architecture, builds upon the U-Net framework, which excels at extracting features from input images to produce outputs that highlight the most relevant information. 
The attention U-Net enhances this by incorporating attention mechanisms to improve feature selection. 
Below, we first introduce the U-Net architecture and then describe the enhancements introduced in the attention U-Net.

The U-Net, originally designed for image segmentation in medical imaging~\citep{Ronneberger2015}, features a symmetric encoder-decoder structure with skip connections to preserve spatial details. 
The encoder path consists of four levels, each comprising two convolutional blocks. 
Each block includes a $3\times3$ convolutional layer followed by an activation function to introduce non-linearity, with the number of feature channels doubling at each level to capture increasingly abstract features. A max-pooling operation downsamples the spatial dimensions by a factor of 2 after each level. 
After four downsampling cycles, the bottleneck layer applies two convolutional blocks to process the deepest, most abstract features.
The decoder path mirrors the encoder, with four levels that progressively upsample the spatial dimensions while halving the number of feature channels.
At each decoder level, skip connections from the corresponding encoder level are concatenated to preserve fine-grained spatial information, followed by two convolutional blocks. 
A final $1\times1$ convolutional layer produces an output with the same dimensions as the input, enabling precise pixel-wise predictions.
This architecture makes U-Net well-suited for mapping noisy SFT data to denoised CW signal outputs.

The attention U-Net enhances this framework by integrating attention gates before concatenating the skip connections~\citep{Oktay2018}. 
These gates compute attention weights to highlight salient regions in the encoder feature maps while suppressing irrelevant or noisy features. 
Specifically, each attention gate takes two inputs: the encoder feature map at the corresponding level and the upsampled feature map from the coarser decoder level. 
Both inputs undergo a $1\times1$ convolutional layer to align their dimensions, followed by an element-wise addition and an activation function to combine spatial and contextual information. 
A subsequent $1\times1$ convolutional layer with a sigmoid activation generates attention coefficients $\psi_\mathrm{a} \in [0, 1]$, which are multiplied with the encoder feature map to produce a weighted feature map. 
This weighted map is then concatenated with the upsampled decoder features, enhancing the model’s ability to focus on relevant CW signal patterns while attenuating noise-related features.

In this work, we replace the commonly used ReLU activation function with Leaky ReLU to allow a small, non-zero gradient for negative inputs, improving training stability and mitigating the risk of dying neurons. 
The input data, consisting of 4 channels, is processed through the encoder, where the number of channels increases to 64 at the first level and doubles at each subsequent level, reaching 1024 at the bottleneck of the network. 
Additionally, we introduce a scaling factor in the attention gate to learn the desirable influence of salient features, further enhancing the model’s ability to prioritize CW signal patterns. 
To ensure the output values lie within $[-1, 1]$, we add a $\tanh$ activation layer at the final output, aligning with the normalized range of the data.
The number of trainable parameters of the attention U-Net is $\approx$ 35\,M.

\subsection{Loss Function}
Typically, U-Net training employs a mean square error (MSE) loss function, averaging the squared pixel-wise differences between predicted and ground truth values. 
To optimize denoising performance on spectrograms, i.e., preserving the CW signal while minimizing background noise, we define a custom loss function as:
\begin{equation} \label{eq: loss function}
\mathcal{L} = \frac{1}{N_{\mathcal{S}}} \sum_{y_i \in \mathcal{S}} (y_i - \hat{y}_i)^2 + \frac{1}{N_{\mathcal{B}}} \sum_{x_i \in \mathcal{B}} (x_i - \hat{x}_i)^2,
\end{equation}
where $\mathcal{S}$ represents the signal region, $\mathcal{B}$ denotes the background (noise) region, and $N_{\mathcal{S}}$ and $N_{\mathcal{B}}$ are the number of pixels in the signal and noise regions, respectively.
Here, $y_i$ and $x_i$ are the predicted values, while $\hat{y}_i$ and $\hat{x}_i$ are the ground truth (clean signal) values for each pixel in the signal and noise regions, respectively. 
The first term minimizes errors in the signal region, ensuring accurate restoration of the CW signal, while the second term targets noise reduction in the background. 
Since CW signals occupy a narrow trajectory in the SFT data, comprising only a small fraction of pixels, this template-guided loss function prevents training from being dominated by noise reduction at the expense of signal restoration. 
The ground truth is derived from clean, simulated CW signals.

To apply this loss function, we define the signal and noise regions using a thresholding approach. 
For each image, we generate a mask by computing the mean $\mu$ and standard deviation $\sigma$ of the absolute pure signal values across both spatial and channel dimensions.
Pixels where the absolute value exceeds $\mu + \sigma$ are classified as part of the signal region $\mathcal{S}$, while the remaining pixels are assigned to the background region $\mathcal{B}$. 

\subsection{Training Process}
To train the attention U-Net, we used the Adam optimizer \citep{Kingma2017} with a batch size of 8 samples and an initial learning rate of $  10^{-4}  $.
Our training process employed a scheduler in PyTorch \citep{Paszke2019} that halved the learning rate if the detection probability (\(\pdet\), defined in Section~\ref{sec:detectability}) of the validation set did not improve for 70 epochs, stopping training after two such decays. The training data consist of 7,000 noisy signal samples and 7,000 pure noise samples, while a validation set that contains 1,000 noisy signal samples and 1,000 pure noise samples is used to monitor the model performance.
To enhance robustness against unseen data and prevent overfitting to specific noise patterns, we dynamically generated Gaussian noise at each training epoch.

\subsection{Detection Statistic and Detectability} \label{sec:detectability}
Our goal is to use the trained model for signal detection. 
To achieve this, we need a detection statistic that is computed from the model's output to quantify the presence of CW signals. 
We define the detection statistic as the mean of the squared pixel values across all spatial and channel dimensions of the output image.
This statistic represents the average power or intensity of the output image. 
The trained attention U-Net enhances the signal while suppressing background noise, producing higher detection statistic values when a signal is present.

In addition to the loss function, we monitored the training process with \(\pdet\), which measures the fraction of noisy CW signal samples correctly identified by the detection statistic at a fixed false alarm rate (\(\pfa\)). 
Using Eq.~3 from \citep{Joshi2024}, we set \(\pfa\) to 1\% per 50\,mHz.
For our data, spanning a 35.56\,mHz frequency band per image, we calculated $  \pfa = 0.71\%  $.
We selecte the model with the highest \(\pdet\) on the validation set as the final model, since \(\pdet\) is the metric for evaluating the model’s ability to identify CW signals in the presence of noise. 

The implementation of the attention U-Net is available in the project repository at \texttt{cw\_unet}~\footnote{\url{https://github.com/damondmc/cw_unet}}.

\section{Results} \label{sec:result}
We trained five models using data spanning 921,600\,s from two detectors, with parameters given in Table~\ref{tab:parameter_ranges}, except for fixed frequencies at 20\,Hz, 200\,Hz, 500\,Hz, 1000\,Hz, and the entire range of 20--1000\,Hz.
The training data have noise levels of \(\sqrt{S_n}/h_0 = 32, 25, 22, 19, 20\,\mathrm{Hz}^{-1/2}\) respectively. 

\begin{figure}[htp]
\centering
\includegraphics[width=0.45\textwidth]{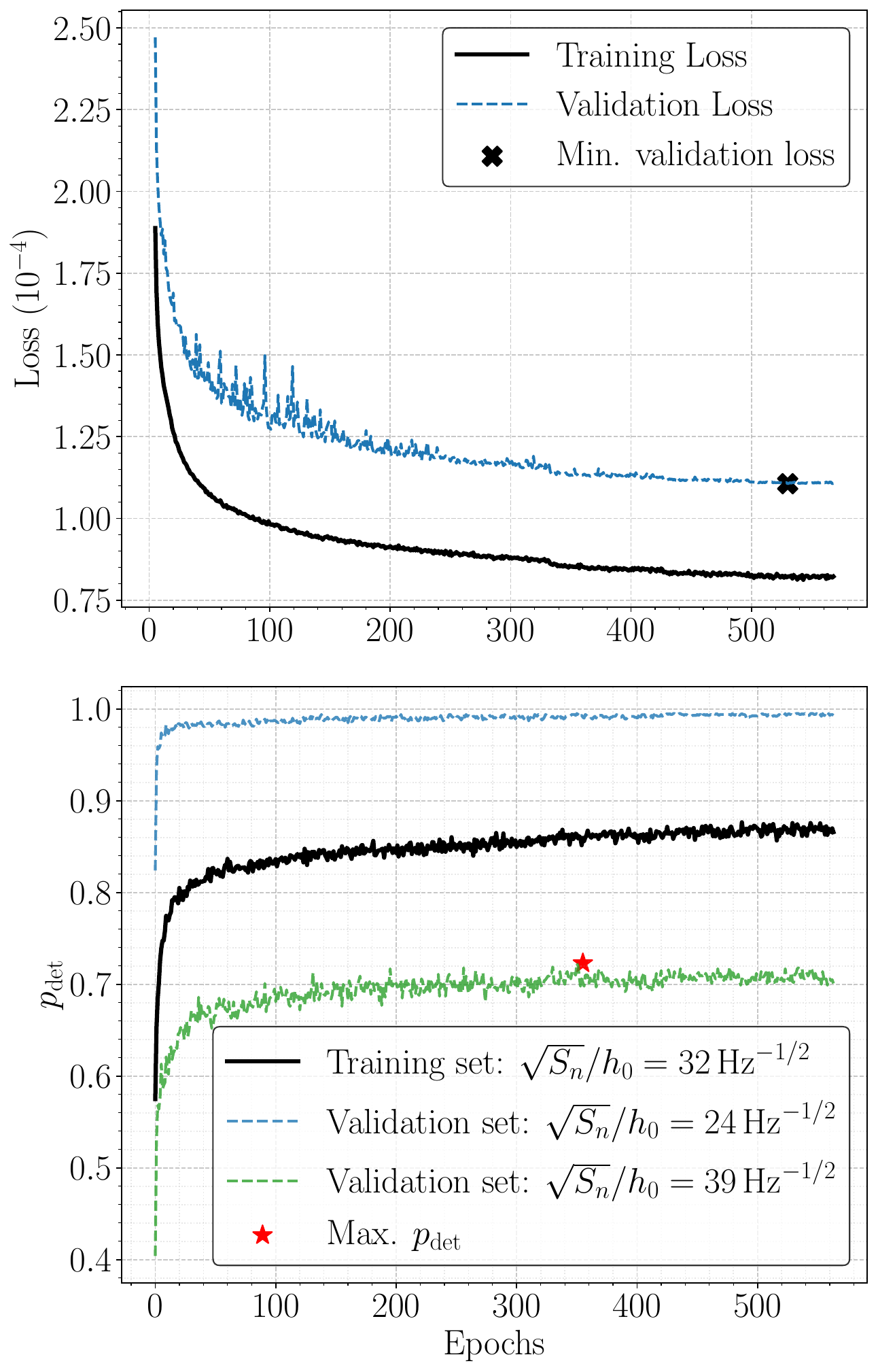}
\caption{Loss (top) and detection probability \(\pdet\) at a \( 1\%\) false alarm rate per \(50\,\mathrm{mHz}\) (bottom) during the training process for the 20\,Hz attention U-Net model on the training dataset (black) and validation dataset (colored) with different noise level \(\sqrt{S_n}/h_0\).}
\label{fig:training pdet}
\end{figure}

The training progression for the 20\,Hz model is shown in Figure.~\ref{fig:training pdet}.
The lowest validation loss and the highest \(\pdet\) for the noisiest validation set (\(\sqrt{S_n}/h_0 = 39\,\mathrm{Hz}^{-1/2}\)) are marked with a black cross and a red star, respectively.
To prioritize sensitivity in CW signal detection, we select the model with the highest \(\pdet\) rather than the lowest validation loss.
The training took $\sim 5$ days using a single NVIDIA Tesla V100 16GB GPU.

\subsection{Denoising} \label{sec:denoising result}

To demonstrate the denoising performance, we tested the trained model on an unseen test dataset, distinct from the training and validation sets. 
The attention U-Net processes the noisy input spectrogram through its encoder-decoder structure, leveraging attention gates to focus on CW signal patterns while suppressing noise, ultimately producing a denoised output that closely resembles the ground truth signal.

\begin{figure*}[htp]
\centering
\includegraphics[width=\textwidth]{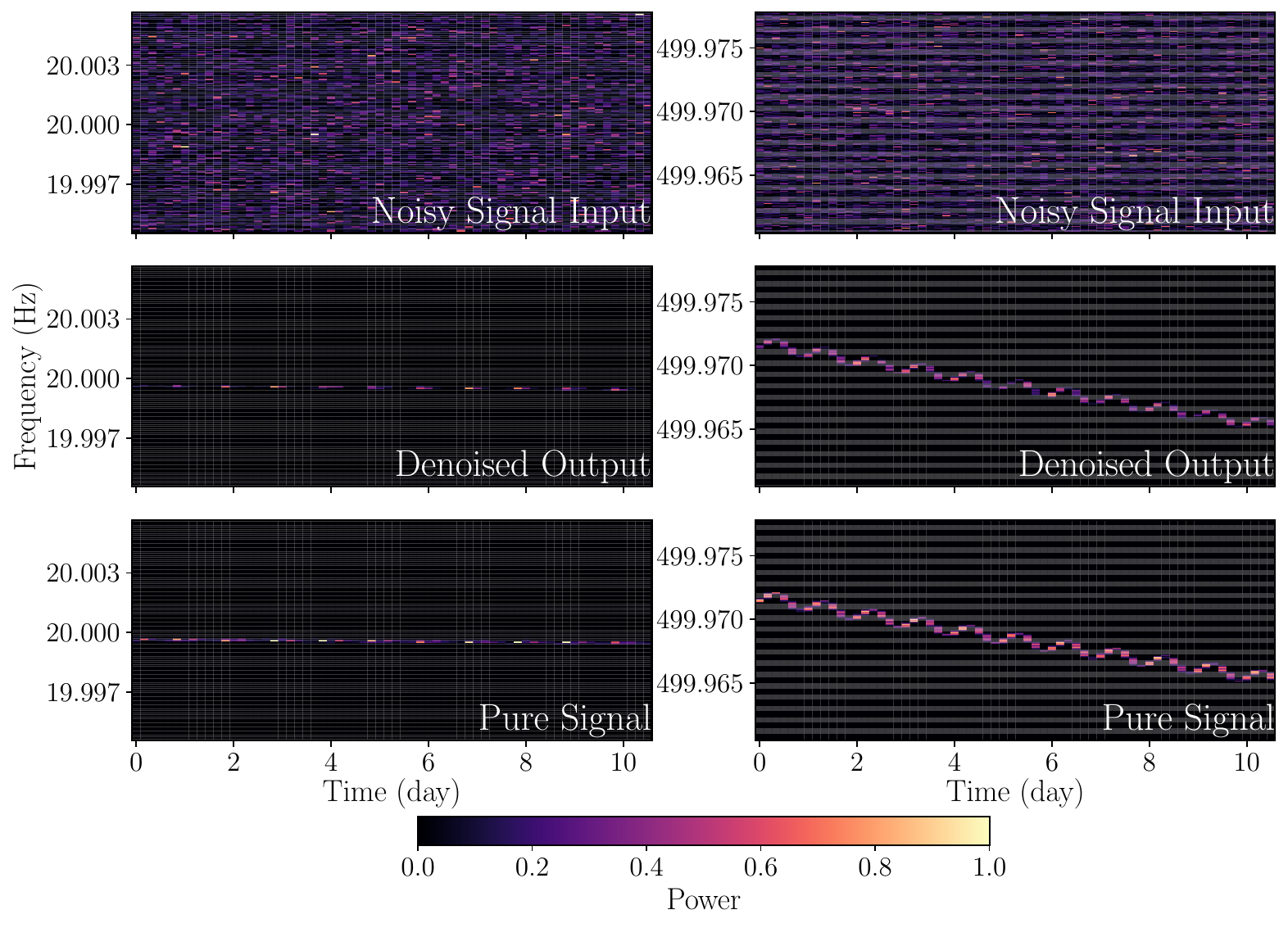}
\caption{Denoising performance of the attention U-Net models trained at 20\,Hz (left) and 500\,Hz (right) on H1 spectrogram spanning 921,600\,s. 
The attention U-Net takes a noisy input with noise level \(\sqrt{S_n}/h_0 = 20\,\mathrm{Hz}^{-1}\) (top) and produces a denoised output (middle), which closely resembles the ground truth signal (bottom), recovering the major signal features.
}
\label{fig:Denoised spectrogram}
\end{figure*}

Figure~\ref{fig:Denoised spectrogram} illustrates an example from the test dataset at 20\,Hz (left) and 500\,Hz (right). 
The top panel shows the noisy input spectrogram with a noise level of \(\sqrt{S_n}/h_0=20\,\mathrm{Hz}^{-1}\).
The middle panel displays the attention U-Net’s denoised output, revealing a clear representation of the signal with significantly reduced noise. 
The bottom panel shows the clean reference signal, which closely matches the output, demonstrating the model’s effectiveness in recovering the underlying CW signal.

\begin{figure*}[htp]
\centering
\includegraphics[width=\textwidth]{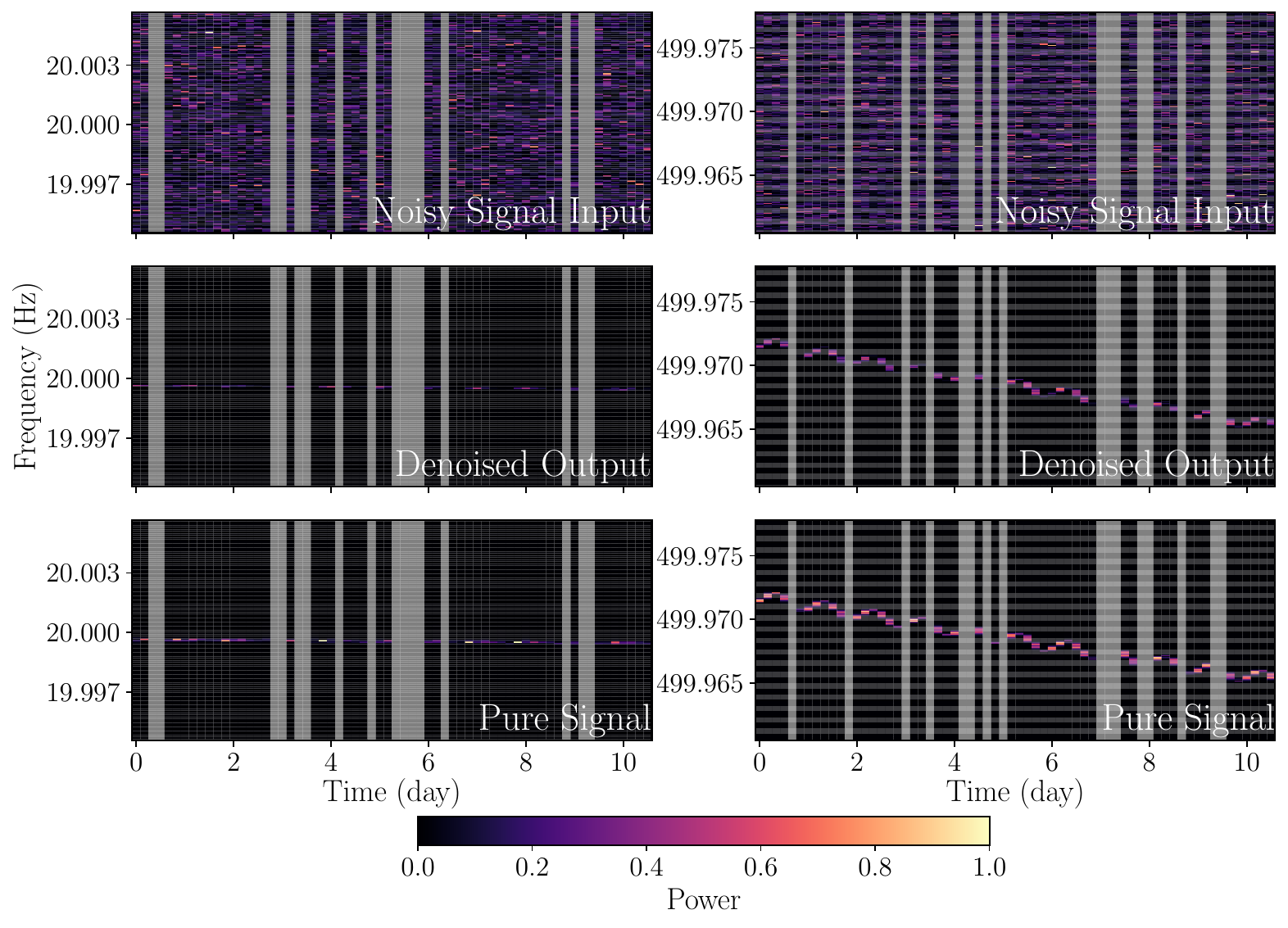}
\caption{Denoising performance of the attention U-Net models trained at 20\,Hz (left) and 500\,Hz (right) on H1 spectrogram spanning 921,600\,s with time gaps (duty factor = 0.75). 
Despite being trained on data without time gaps, the models successfully denoise the noisy input with noise level \(\sqrt{S_n}/h_0 = 20\,\mathrm{Hz}^{-1}\) (top) to produce an output (middle) that closely resembles the ground truth signal (bottom), recovering the major signal features.}
\label{fig:Denoised spectrogram with gaps}
\end{figure*}

We also evaluated the model’s robustness on test data with time gaps, simulating real-world conditions where ground-based interferometers experience offline periods during observation runs. 
Figure~\ref{fig:Denoised spectrogram with gaps} shows the denoising result for an input with a duty factor of 0.75, meaning 25\% of the observation time contains no data.
Despite the absence of time gaps in the training set, the attention U-Net successfully recovers the major features of the CW signal, demonstrating its generalizability to realistic data conditions.

\subsection{Sensitivity} \label{sec:sensitivity}
To assess the model’s sensitivity, it's useful to introduce a figure of merit, the sensitivity depth $\mathcal{D}$~\citep{Behnke2015}:
\begin{equation} \label{eq:depth}
    \depth \equiv \frac{\sqrt{S_n}}{h_0},
\end{equation}
which represents the sensitivity level at which a search can detect a signal with an upper limit amplitude $h_0$ in data with noise strain amplitude $\sqrt{S_n}$. 
A higher sensitivity depth indicates a more sensitive search. Here, we use the 90\% upper limit on the signal strain amplitude and denote the corresponding sensitivity depth as $\D90$.

\begin{figure}[htp]
\centering
\includegraphics[width=0.45\textwidth]{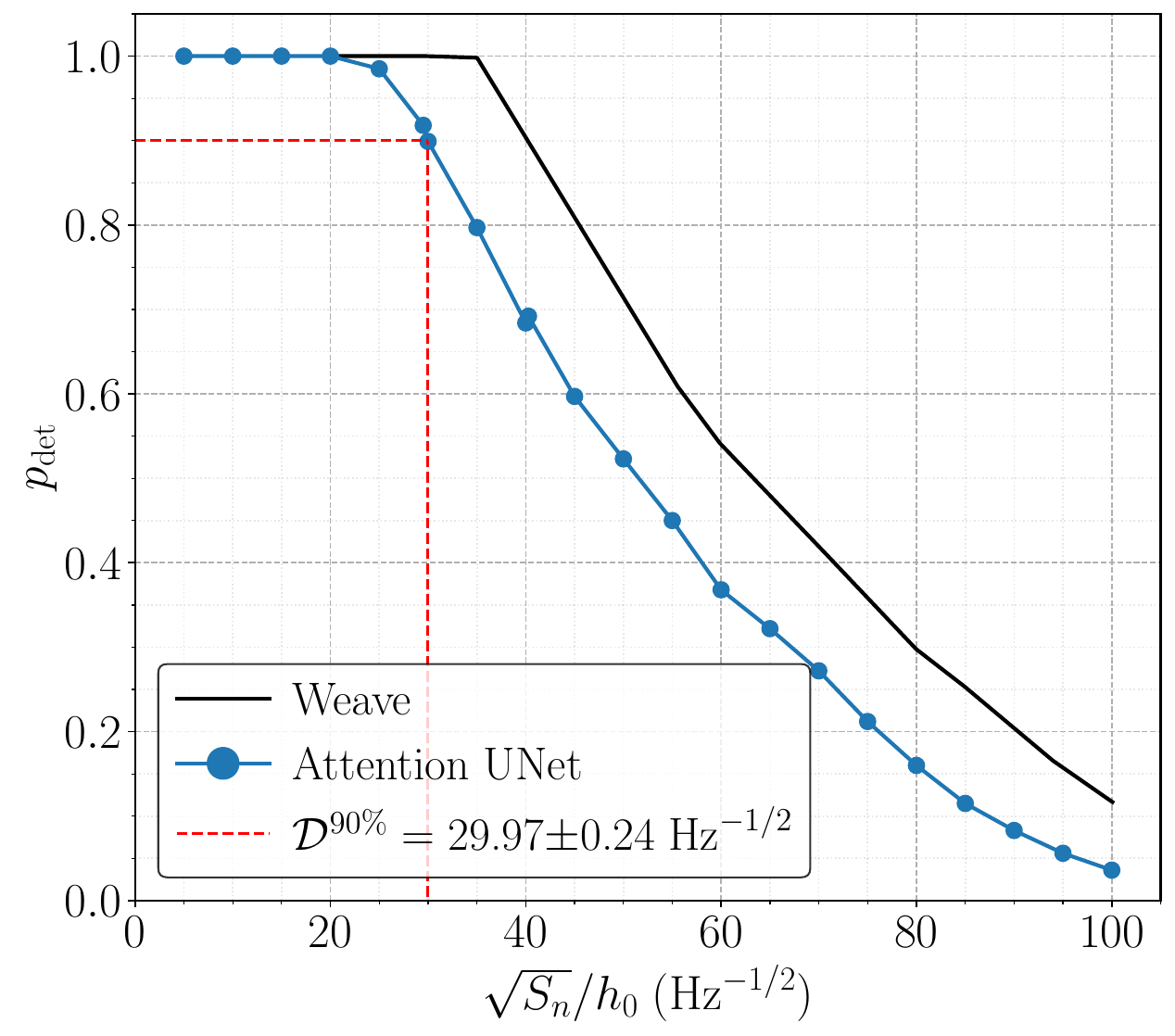}
\caption{
Detection probability \(\pdet\) (at a \(1\%\) false alarm rate per \(50\,\mathrm{mHz}\)) versus noise level \(\sqrt{S_n}/h_0\) at 20\,Hz. 
The black curve shows the performance of the Weave matched-filter method, rescaled from~\citep{Joshi2024}.
The blue curve represents the attention U-Net's performance, with a 90\% sensitivity depth \(\D90 = 29.97\pm 0.24\,\mathrm{Hz}^{-1/2}\).}
\label{fig:depth 20Hz}
\end{figure}

We evaluated \(\pdet\) across various noise levels \(\sqrt{S_n}/h_0\) using 2,000 test samples (1,000 noisy signals and 1,000 pure noise) to determine the noise level at which a 90\% detection probability is achieved (\(\D90\)). 
Figure~\ref{fig:depth 20Hz} shows the \(\pdet\) of the 20\,Hz model at different noise levels (blue curve).
We also included the \(\pdet\) of a matched-filter search using Weave from~\citep{Joshi2024}, scaled to match the timespans in this study, as a benchmark (black curve).
Our model achieves a sensitivity depth of \(\D90 = 29.97 \pm 0.24\,\mathrm{Hz}^{-1/2}\).
To compare with~\citep{Joshi2024}, we scaled their \(\D90\) values using the relation \(\D90\propto\Tobs^{0.25}\), adjusting from their 10-day observation period to our 921,600\,s ($\approx$ 10.67 days). 
This yields a sensitivity depth of \(29.78\,\mathrm{Hz}^{-1/2}\) for their model.
The sensitivity of our attention U-Net is comparable to the model from~\citep{Joshi2024} trained at 20\,Hz.

The performance of all models is summarized in Table~\ref{tab:sensitivity}, which reports the sensitivity depth \(\D90\) of our trained model, the rescaled sensitivity depth using the Weave matched-filter method $\depthmf$ from~\citep{Joshi2024}, and the detectability of our trained model at a noise level $\sqrt{S_n}/h_0=\depthmf$.
Our attention U-Net achieves comparable sensitivity at 20\,Hz and 200\,Hz but lower sensitivity at 500\,Hz, 1000\,Hz, and acroos the 20--1000\,Hz range compared to~\citep{Joshi2024}.

\begin{table}[htp]
\centering
\begin{tabular}{lccc}
\hline 
Frequency & \(\mathcal{D}^{90\%}\,[\mathrm{Hz}^{-1/2}]\) & \(\mathcal{D}_\mathrm{MF}^{90\%}\,[\mathrm{Hz}^{-1/2}]\) & \(p_\mathrm{det}\) \\
\hline\hline
20\,Hz & \(29.97\pm 0.24\) & 40.28 & \(69.2 \pm 1.5\)\% \\
200\,Hz & \(22.41\pm 0.18\) & 37.80 & \(46.4 \pm 1.6\)\% \\
500\,Hz & \(22.08\pm 0.16\) & 36.77 & \(40.1 \pm 1.5\)\% \\
1000\,Hz &\(17.10\pm 0.12\)& 34.50 & \(32.8 \pm 1.5\)\% \\
20--1000\,Hz & \(18.63\pm 0.24\) & -- & \(33.6 \pm 1.5\)\% \\
\hline
\end{tabular}
\caption{Sensitivity depth at 90\% detection probability (\(\mathcal{D}^{90\%}\)) for the trained attention U-Net models, compared to the matched-filter sensitivity depth (\(\D90_\mathrm{MF}\)) from~\citep{Joshi2024}, and the detection probability of the trained attention U-Net at noise level $\sqrt{S_n}/h_0=\mathcal{D}_\mathrm{MF}^{90\%}$.}
\label{tab:sensitivity}
\end{table}

\subsection{Generalization on Duty Factor and Total Observation Time} \label{sec:generalization}

\begin{figure}[htp]
\centering
\includegraphics[width=0.45\textwidth]{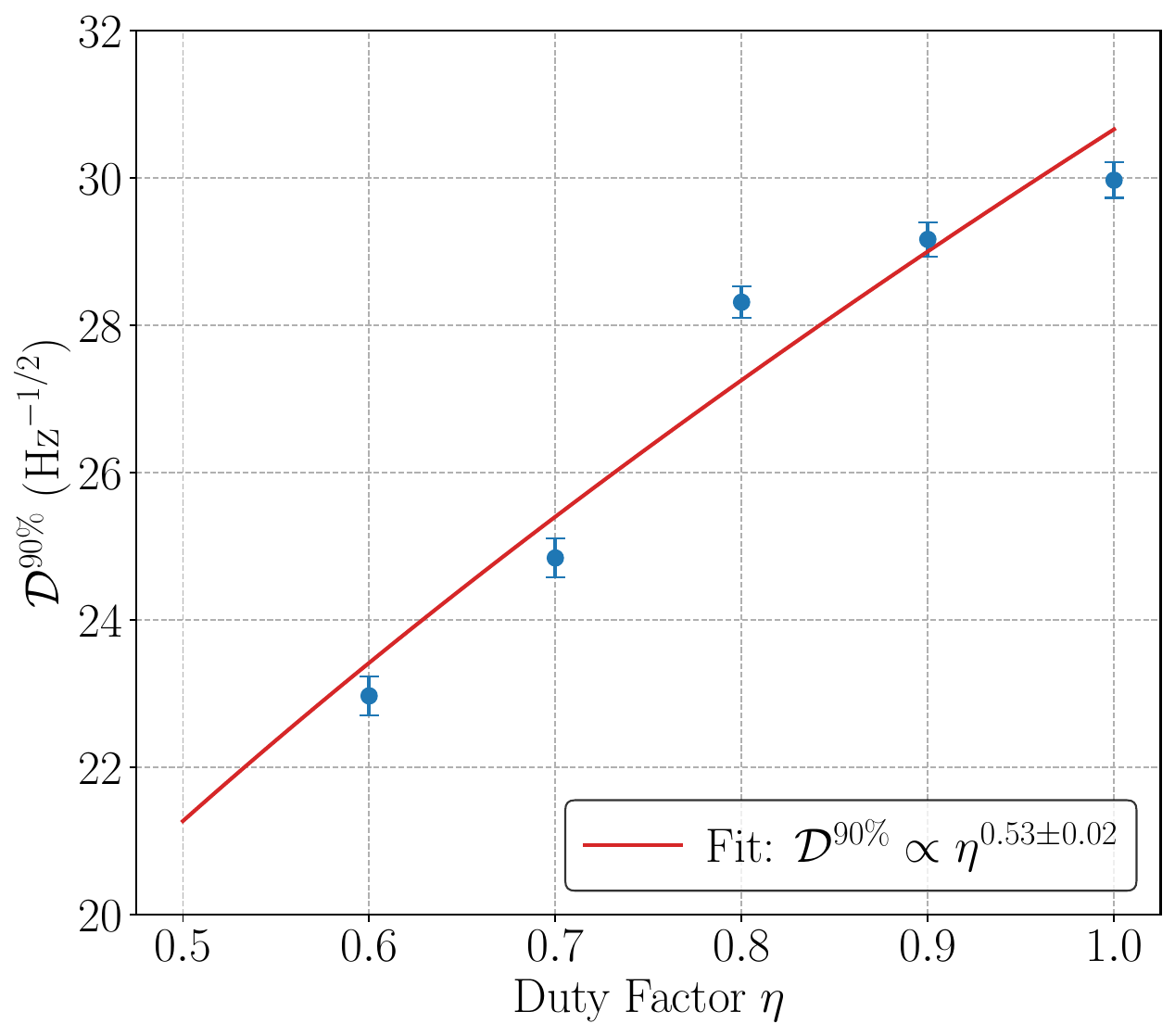}
\caption{Sensitivity depth \(\D90\) versus duty factor \(\eta\) for the 20\,Hz attention U-Net model trained on gap-free data.
For $\eta = 0.6$, the model achieves a sensitivity depth of \(\D90= 22.97\pm 0.26\,\mathrm{Hz}^{-1/2}\), with a fitted power law indicating that \(\D90 \propto \eta^{0.53 \pm 0.02}\).}
\label{fig:depth vs duty factor}
\end{figure}

To study the dependence of sensitivity on the duty factor \(\eta\), we evaluated the performance of the 20\,Hz model on test data with time gaps, simulating realistic observation conditions. 
Figure~\ref{fig:depth vs duty factor} shows the variation of \(\D90\) with duty factors ranging from 0.5 to 1.0, where typical observation runs have duty factors of approximately 0.6--0.7. 
Fitting \(\D90\) to a power law reveals a scaling of \(\D90 \propto \eta^{0.53 \pm 0.02}\), with \(\D90= 22.97\pm 0.26\,\mathrm{Hz}^{-1/2}\) at \(\eta = 0.6\).

\begin{figure}[htp]
\centering
\includegraphics[width=0.45\textwidth]{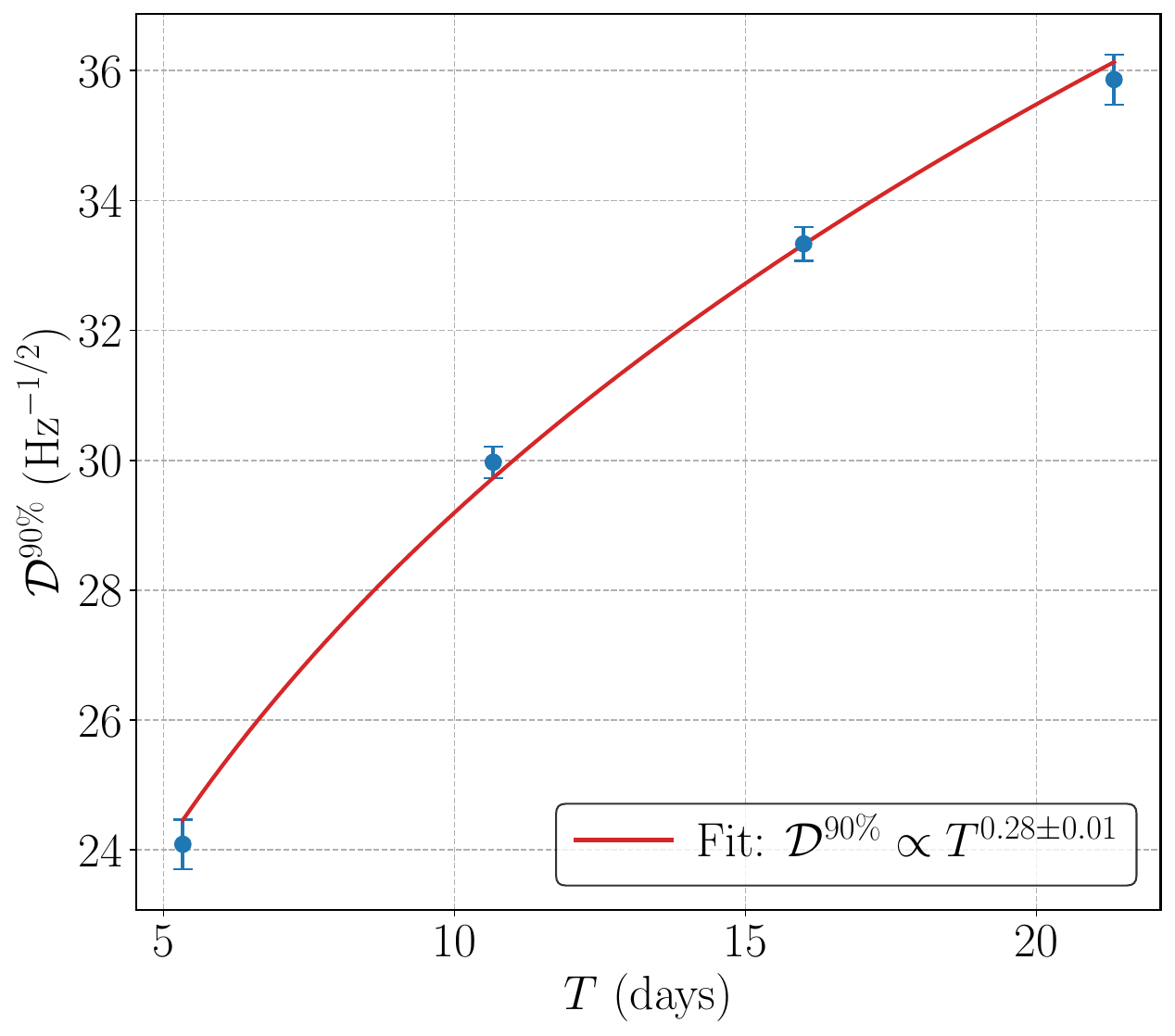}
\caption{Sensitivity depth \(\D90\) versus total observation time \(\Tobs\) for the 20\,Hz attention U-Net model. 
For \(\Tobs = 1,832,000\,\mathrm{s}\) (\(\approx\) 21 days), the model achieves a sensitivity depth of \(\D90 = 35.86\pm 0.38\,\mathrm{Hz}^{-1/2}\), with a fitted power law indicating that \(\D90 \propto \Tobs^{0.28 \pm 0.01}\).}
\label{fig:depth vs tobs}
\end{figure}

Our neural network was initially trained on data with \(\Tobs \approx\) 10.67 days.
To maximize sensitivity, we aim to use as much data as possible.
Therefore, we investigate the impact of \(\Tobs\) on the sensitivity by training the 20\,Hz model with $\Tobs=$ 460,800\,s, 921,600\,s, 1,382,400\,s, and 1,832,000\,s (approximately 5, 10, 16, and 21 days, respectively). 
Figure~\ref{fig:depth vs tobs} shows the resulting \(\D90\) values, achieving \(\D90 = 35.86\pm 0.38\,\mathrm{Hz}^{-1/2}\) at \(\Tobs \approx\) 21 days. 
The sensitivity depth scales as \(\D90 \propto \Tobs^{0.28 \pm 0.01}\), similar to the \(\Tobs^{0.25}\) scaling observed in semi-coherent search methods. 
Notably, the training time for our attention U-Net scales linearly with $\Tobs$, as the number of frequency bins is fixed.
For sufficiently long \(\Tobs\), such as a year of data, we have to increase the number of time bins by a factor of 35 and frequency bins by a factor of 7 to cover the entire signal for the parameter range defined in Table.~\ref{tab:parameter_ranges}.
As a result, the overall scaling remains lower than that of semi-coherent searches, which scale as \(\sim \Tobs^2\) for a fixed coherence time~\citep{Prix2012}.  
This demonstrates that the attention U-Net architecture is feasible to scale up to longer observation time for better performance.

\subsection{Sky position} 

In this section, we investigate the sensitivity of the all-sky search as a function of the CW source’s sky position. 
To obtain the intrinsic dependence on sky position, we fixed the SNR of the injected signal by adjusting the noise level, thereby removing the influence of detector antenna patterns~\citep{Dreissigacker2020}, as given by:
\begin{equation} \label{eq:geometric factor}
\rho^2 = \frac{4}{25} \frac{T}{D^2} R^2(\theta),
\end{equation}
where \( R^2(\theta) \) is a geometric factor that depends on \( \{\alpha, \delta, \cos \iota, \psi\} \).
By fixing the SNR to achieve an overall \(\pdet\approx 50\%\) across all sky positions, we calculate the overall \(\pdet\) for various sky positions using the noise level derived from Eq.~\ref{eq:geometric factor}.
\begin{figure*}[htp]
\centering
\includegraphics[width=\textwidth]{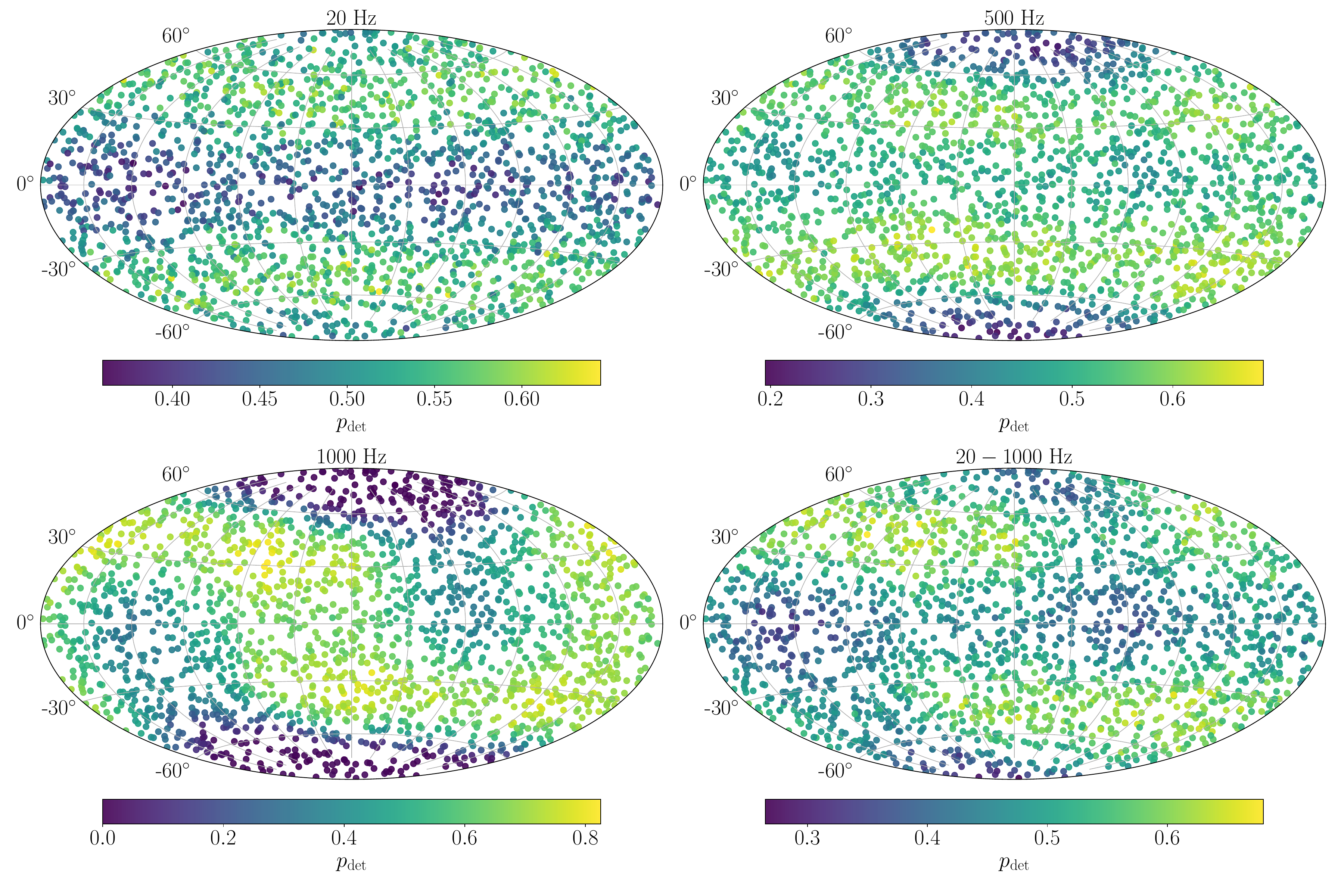}
\caption{Detection probability $\pdet$ (at a $1\%$ false alarm rate per $50\,\mathrm{mHz}$) of the attention U-Net models as a function of the CW source sky position (Hammer projection).
The SNR is fixed to have an overall $\pdet \approx 50\%$.
The top panel shows the all-sky detectability for models trained at 20 Hz (left) and 500 Hz (right).
The bottom panel shows the all-sky detectability for models trained at 1000 Hz (left) and 20--1000 Hz (right).
}
\label{fig:sensitivity skymap}
\end{figure*}

In Figure~\ref{fig:sensitivity skymap}, we plot $\pdet$ as a function of sky position for different trained models. 
In the top panel, the $20\,\mathrm{Hz}$ model (top left) exhibits a clear pattern of lower $\pdet$ near the celestial equator and higher $\pdet$ for signals originating near the celestial poles. 
Similarly, the $500\,\mathrm{Hz}$ model (top right) shows a comparable pattern but with relatively lower $\pdet$ near the celestial poles. 
In contrast, the $1000\,\mathrm{Hz}$ model (bottom left) shows an irregular sensitivity pattern with no clear directional preference. 
The model trained on the $20$--$1000,\mathrm{Hz}$ band (bottom right) displays a mixture of these sensitivity patterns. 
\citep{Dreissigacker2020} and \citep{Joshi2023} pointed out CNN-based neural network struggles to learn the structure of a comparatively wider signal due to their larger Doppler modulation.
Our results align with this, indicating that the attention U-Net is less sensitive to signals coming from the celestial equator ($\delta \approx 0$), where Doppler modulation is larger, and at higher frequencies, where the signal’s frequency span increases proportionally.

\section{Discussion} \label{sec:discussions}

The attention U-Net used in this work effectively denoises and detects CW signals in noisy data, leveraging a distinct architecture compared to existing methods.
By employing a normalized loss function (Eq.~\ref{eq: loss function}) that balances signal restoration and noise suppression, along with a detection statistic based on mean power across image dimensions, the model successfully recovers CW signals, as shown in Figures~\ref{fig:Denoised spectrogram} and \ref{fig:Denoised spectrogram with gaps}.
The attention U-Net shows comparable sensitivity to that reported in~\citep{Joshi2024}.
While~\citep{Joshi2024} offers a better high-frequency sensitivity, our architecture balances performance and efficiency as a promising alternative.
The model demonstrates robustness to data time gaps, with sensitivity $\propto \eta^{0.53 \pm 0.02}$ with duty factor. 
The model's sensitivity scales with the total timespans of the data as $\propto \Tobs^{0.28 \pm 0.01}$, while the training time scales as $\sim T^m$, where $1 < m < 2$.
Sky position analysis from~Figure~\ref{fig:sensitivity skymap} reveals frequency-dependent sensitivity variations, with higher detectability away from the celestial equator at lower frequencies and more irregular sensitivity patterns at higher frequencies, indicating challenges due to broader Doppler modulations.

To enhance performance, we attempted curriculum learning by starting with less noisy data and gradually increasing the noise level, but it yielded no clear sensitivity gains.
However, scaling up to 20,000 training samples (10,000 signals + 10,000 pure noise) and 128 initial feature channels improved $\D90$ to $18.00\pm 0.12\,\mathrm{Hz}^{-1/2}$ at 1000\,Hz, suggesting optimization potential.
Due to limited resources, we show the results as a proof-of-concept, showing U-Net-based architectures are viable for all-sky CW searches over wide parameter spaces, rather than to train the model with more data or channels.
For addressing Doppler modulation challenges, increasing convolutional kernel sizes (e.g., to 4$\times$4 or 5$\times$5) was tested but led to unstable training and reduced sensitivity, warranting further investigation into effective capture of larger signal patterns.
Similarly, integrating residual blocks—as successfully applied in~\citep{Dreissigacker2020} and~\citep{Joshi2024}—through a Residual U-Net~\citep{Alom2018} was attempted, but resulted in unstable training and lower sensitivity compared to the attention U-Net.
Additional studies are needed to study the reason behind it.

Looking ahead, future work could include fine-tuning high-frequency models with larger datasets, incorporate time gaps in training for greater robustness, or replace MaxPooling with wavelet transforms~\citep{Li2020} to better preserve weak signals.
Evaluating non-stationary noise in long observations is also essential for real data applications.
The neural network could be applied to all-sky searches for CWs emitted from binary NSs by training on data preprocessed using the TwoSpect method~\citep{Goetz2011}, which transforms long-duration signal spectrograms into localized patterns that may be less affected by the kernel size limitations encountered here.
The attention U-Net’s output could also serve as input to a variational autoencoder~\citep{Gabbard2021} for rapid parameter estimation.
Finally, having demonstrated the capability of the U-Net architecture to detect CW signals, exploring diffusion models~\citep{Ho2020,Song2021,Legin2025}—advanced denoising frameworks that employ U-Net as a backbone—may further enhance the sensitivity of machine learning–based CW searches.

\section{Acknowledgments}
The author thanks K. Riles, R. Prix, and A. Rodríguez for helpful discussion and suggestions. 
The author also thanks K. Riles and T. S. Yamamoto for helpful comments on this manuscript. 
This research was supported in part by National Science Foundation Award PHY-2408883 and through computational resources and services provided by Advanced Research Computing at the University of Michigan, Ann Arbor.

\newpage 

\bibliography{bibliography}

\label{lastpage}
\end{document}